\documentclass[sigconf,nonacm]{acmart}

\usepackage{graphicx}
\newcommand{\smartparagraph}[1]{\vspace{.05in}\noindent\textbf{#1}}
\usepackage[colorinlistoftodos,prependcaption]{todonotes}
\usepackage{listings}
\usepackage{xspace}
\usepackage{tabularx}
\usepackage{multirow}
\usepackage{makecell}
\usepackage{dcolumn}
\usepackage{multirow}
\usepackage{paralist}
\usepackage{enumitem}
\usepackage{booktabs}
\usepackage{multirow}
\usepackage[most]{tcolorbox}
\usepackage{dafny_colors}
\usepackage{xurl}
\def\BibTeX{{\rm B\kern-.05em{\sc i\kern-.025em b}\kern-.08em
    T\kern-.1667em\lower.7ex\hbox{E}\kern-.125emX}}
    
\AfterEndEnvironment{minted}{\vspace{-15pt}}

\usepackage{tcolorbox}

\setlist{nolistsep}

\makeatletter
\newcommand{\DeclareLatinAbbrev}[2]{%
  \DeclareRobustCommand{#1}{%
    \@ifnextchar{.}{\textit{#2}}{%
      \@ifnextchar{,}{\textit{#2.}}{%
        \@ifnextchar{!}{\textit{#2.}}{%
          \@ifnextchar{?}{\textit{#2.}}{%
            \@ifnextchar{)}{\textit{#2.}}{%
              {\textit{#2.,\ }}}}}}}}%
}
\makeatother
\DeclareLatinAbbrev{\eg}{e.g}
\DeclareLatinAbbrev{\Eg}{E.g}
\DeclareLatinAbbrev{\ie}{i.e}
\DeclareLatinAbbrev{\Ie}{I.e}
\DeclareLatinAbbrev{\etc}{etc}
\DeclareLatinAbbrev{\etal}{et~al}

\def\first {$(i)$\xspace}
\def\second{$(ii)$\xspace}
\def\third {$(iii)$\xspace}

\newcommand{\systemname}{\textsc{Prometheus}\xspace}

\newcommand*\circled[1]{\tikz[baseline=(char.base)]{
            \node[shape=circle,draw,inner sep=1pt] (char) {#1};}}

\usepackage{etoolbox} 

\makeatletter
\patchcmd{\lst@makecaption}
  {\@tempdima\hsize}
  {\@tempdima\linewidth}
  {}{}
\renewcommand{\lst@makecaption}[2]{%
  \vskip\abovecaptionskip
  \begin{center}
    \normalfont \footnotesize #1: #2
  \end{center}
  \vskip\belowcaptionskip
}
\makeatother

\begin{document}

\title[Dissect-and-Restore: AI-based Code Verification with Transient Refactoring]{Dissect-and-Restore: \\ AI-based Code  Verification with Transient Refactoring}



\author{Changjie Wang}
\affiliation{
  \institution{KTH Royal Institute of Technology}
  \city{Stockholm}
  \country{Sweden}
}
\email{changjie@kth.se}

\author{Mariano Scazzariello}
\affiliation{
  \institution{RISE Research Institutes of Sweden}
  \city{Stockholm}
  \country{Sweden}
}
\email{mariano.scazzariello@ri.se}

\author{Anoud Alshnakat}
\affiliation{
  \institution{KTH Royal Institute of Technology}
  \city{Stockholm}
  \country{Sweden}
}
\email{anoud@kth.se}

\author{Roberto Guanciale}
\affiliation{
  \institution{KTH Royal Institute of Technology}
  \city{Stockholm}
  \country{Sweden}
}
\email{robertog@kth.se}

\author{Dejan Kosti\'c}
\affiliation{
  \institution{KTH Royal Institute of Technology}
  \city{Stockholm}
  \country{Sweden}
}
\email{dmk@kth.se}

\author{Marco Chiesa}
\affiliation{
  \institution{KTH Royal Institute of Technology}
  \city{Stockholm}
  \country{Sweden}
}
\email{mchiesa@kth.se}

\renewcommand{\shortauthors}{Trovato et al.}

\newcommand{\aachange}[2]{ {\color{red}{#1}} {\color{blue}{#2}}}
\newcommand{\postdeadlinefixed}[1]{ {\color{red}{#1}}}

\begin{abstract}
Formal verification is increasingly recognized as a critical foundation for building reliable software systems. However, 
the need for specialized expertise to write precise specifications, navigate complex proof obligations, and learn annotations, often makes verification order of magnitude more expensive than implementation. While modern AI systems can recognize patterns in mathematical proofs and interpret natural language, effectively integrating them into the formal verification process remains an open challenge.

We present \systemname, a novel AI-assisted system that facilitates automated code verification with current AI capabilities in conjunction with modular software engineering principles (\eg modular refactoring).
Our approach begins by decomposing complex program logic, such as nested loops, into smaller, verifiable components. Once verified, these components are 
recomposed to construct a proof of the original program.
This decomposition-recomposition workflow is non-trivial. 
\systemname addresses this by guiding the proof search through structured decomposition of complex lemmas into smaller, verifiable sub-lemmas. When automated tools are insufficient, users can provide lightweight natural language guidance to steer the proof process effectively.

Our evaluation demonstrates that transiently applying modular restructuring to the code 
substantially improves the AI's effectiveness in verifying individual components. 
This approach successfully verifies 86\% of tasks in our curated dataset, compared to 68\% for the baseline. Gains are more pronounced with increasing specification complexity, improving from 30\% to 69\%, and when integrating proof outlines for complex programs, from 25\% to 87\%.

\end{abstract}




\keywords{Verification Automation, Refactoring, Large Language Models, Dafny}


\maketitle

\section{Introduction}




AI-powered code assistants are revolutionizing the software development landscape. Tools like GitHub Copilot~\cite{copilot}, Cursor AI~\cite{cursor}, and Amazon Q Developer~\cite{amazonQ} rapidly transform how developers build, maintain, and improve software by offering code autocompletion suggestions, automated refactoring, and even natural-language-to-code translation. 
Despite the productivity gains enabled by AI-powered code assistants, their limitations are becoming increasingly apparent. One particularly serious concern is the phenomenon of hallucinations, \ie plausible but incorrect outputs~\cite{sarkar2024copilot}. 
These errors can extend development time and, if not caught during review, may introduce significant security, reliability, or correctness vulnerabilities into the codebase~\cite{spracklen2025packageyoucomprehensiveanalysis,10.1145/3610721}.

One promising way to guard against AI hallucinations in code generation is through \textit{formal verification}, which checks all possible behaviors of a program to ensure correctness. Unlike testing or heuristic-based validation, formal methods can offer strong guarantees that eliminate entire classes of errors, including those introduced by AI-generated artifacts. Importantly, formal verification has long been a cornerstone of high-assurance software development, well before the advent of AI-powered code assistants. Industry leaders have integrated formal methods into production workflows to ensure safety, security, and correctness. For example, Amazon uses Dafny to verify critical AWS authentication services~\cite{dafny_aws_auth}, and Microsoft employs F* to prove the end-to-end security of cryptographic libraries~\cite{zinzindohoue2017hacl}. 


Despite growing industrial interest, formal verification remains difficult to apply broadly. The limited adoption of formal verification is largely due to the steep learning curve of verification tools and the expertise required to use them effectively.
Even for experts, translating intuitive reasoning about correctness into machine-checkable proofs is a time-consuming and meticulous process. Every step must be formally justified, including those that may seem obvious to a human reader.
Even highly automated and user-friendly tools like Dafny~\cite{leino2010dafny} still require non-trivial annotations and auxiliary lemmas to succeed. As a result, only organizations with dedicated formal methods teams, typically some of the large tech companies, can afford the sustained investment needed to verify and maintain correctness as codebases evolve over time.

This is where Large Language Models (LLMs), the foundation of modern AI code assistants, offer a promising shift in how we approach formal verification. The same LLM that produces code, whether correct or potentially hallucinated, could 
also generate the verification code needed to establish its correctness. Given a formal specification, the model can produce assertions, invariants, and auxiliary lemmas to construct a formal proof. 
LLMs significantly reduce the manual burden of formal verification, making them powerful tools for both improving the reliability of AI-generated code and lowering the overall cost of applying formal methods.  

Unfortunately, state-of-the-art LLM-based systems for verifying code can successfully reason only about simple programs that \eg perform basic operations in a single loop~\cite{misu2024towards,sun2024clover,mirchev2024assured,parnian2024pccc}. Based on our experiments, LLMs struggle to structure and refine complex verification proofs, lacking the true reasoning capabilities that allow them to incorporate feedback provided by a verifier correctly.  

%

In this work, we introduce \systemname, a system that facilitates AI-based code verification by leveraging \textit{fundamental} software engineering principles. The key insight behind \systemname is that complex programs can be made more amenable to AI-based automated reasoning by first transforming them into semantically equivalent, modular components. These transformations, \eg function extraction or control flow simplification, decompose the original code into smaller, more manageable units that are easier for AI systems to verify. Once these components are individually verified, \systemname systematically translates them back to the original code structure. 

Modularity offers two main advantages. First, it increases the speed and likelihood of successfully verifying the code in a simplified, refactored form, since smaller components are more tractable for current AI models. Second, once this verified version exists, even if the code structure differs from the original, it serves as a solid \textit{proof anchor} that significantly reduces the risk of AI following incorrect or speculative proof strategies. This early grounding helps constrain the proof search space and guides the AI more reliably as it transforms the proof back to the original code. In doing so, \systemname mitigates common failure modes of AI-driven verification, making the overall process more tractable, robust, and~efficient.

While modularization improves tractability, it does not eliminate all challenges. Some components still require auxiliary lemmas or invariants that are not immediately evident. A central limitation of current AI models is their inability to recognize when a proof goal is fundamentally unprovable, often resulting in wasted effort on infeasible paths. \systemname mitigates this by guiding proof search toward promising directions and away from dead ends. It incrementally decomposes complex conditions into smaller, verifiable sub-lemmas and uses feedback from failed attempts to refine its strategy. To support this process, \systemname includes an ad-hoc feedback mechanism that helps identify appropriate proof techniques, such as adding small surgical assertions, more complex inductions, or strengthening invariants, based on the nature of the failed goal. When needed, \systemname also incorporates the verification process with natural language hints from users to further steer the process, avoiding the need for full formal annotations.

\smartparagraph{Contributions.} In this paper, we make the following contributions:
\begin{itemize}[leftmargin=*,noitemsep]
    \item We characterize the challenges in generating verification proofs brought by various formal specification definitions, specific quirks of verifiers, and code complexities.
    \item We present \systemname, the first system capable of overcoming the limited reasoning and scalability capabilities of AI-based system by refactoring the code into smaller parts and distilling the obtained proof of correctness back into the original code.
    \item We produce an advanced code verification benchmark derived from non-trivial algorithmic questions.
    \item We show that \systemname can solve all the tasks in state-of-the-art benchmarks, and improve success rate from 68\% to 86\% in our curated dataset of non-trivial verification tasks. The improvements are more substantial with complex specifications, rising from 30\% to 69\%, and with the use of proof outlines for challenging programs, reaching 87\% from a baseline of 25\%.
\end{itemize}
\section{Background and Running Example}\label{sec:motivation}


In this section, we provide background on formal verification and we highlight common challenges in formal code verification, addressing difficulties faced by both human developers and AI-based systems.
To illustrate these challenges, we use the Dafny programming language~\cite{dafny-reference-manual} and the \texttt{MaxSub} problem as a case study.
We particularly focus on Dafny, rather than systems like F*~\cite{f_star} or Lean~\cite{lean}, due to its strong emphasis on automation rather than tactics and explicit reasoning. 



\smartparagraph{The \texttt{MaxSub} problem.} Given a sequence of integers \texttt{ints}, compute the \textit{maximum sum} of the elements in any \textit{contiguous} subsequence of \texttt{ints}. The sum of an empty sequence is zero.
\vspace{.05in}

\smartparagraph{A simple implementation.} Implementing a correct solution for \texttt{MaxSub} is trivial, as shown in Listing~\ref{code:maxSubCodeImpl}.
The \texttt{MaxSubImpl} algorithm iterates over all possible subsequences and keeps track of the subsequence with the maximum sum. More specifically, for each starting index \texttt{start} (line 4), it slices the remaining part of the array into \texttt{slice} (line 6), and then iterates over all ending indices \texttt{end} in \texttt{slice} to accumulate the sum of elements (lines 7-8). At each step, it updates \texttt{maxSum} if the sum of the current contiguous subsequence is greater than the existing maximum sum (line 9). Finally, it returns the largest sum found (line 12).\footnote{\texttt{ints[start..end]} denotes the subsequence of \texttt{ints} starting between index \texttt{start} (included) and index \texttt{end} (excluded). Index \texttt{end} can be omitted to take the remaining of the sequences as in \texttt{nums[start..]}. \texttt{|ints|} denotes the cardinality of the sequence. }

\begin{lstlisting}[language=dafny, caption={Code implemention for the \texttt{maxSub} problem.}
, label={code:maxSubCodeImpl}, firstnumber=1,xleftmargin=15pt,numbersep=5pt]
method MaxSubImpl(ints: seq<int>) returns(maxSum: int)
ensures IsMaxSubSum(ints, maxSum)
{
  maxSum := 0;  
  for start := 0 to |ints| {
    curr := 0;  
    slice := ints[start..];
    for end := 0 to |slice| {           
      curr := curr + slice[end];
      maxSum := if curr > maxSum then curr else maxSum
    }   
  }
  return maxSum;
}
\end{lstlisting}


\smartparagraph{Verifying correctness: the formal specification.} We now want to verify that the above algorithm returns the correct 
solution for the \texttt{MaxSub} problem. To formally verify that the algorithm is correct we need a formal specification of what the code should compute.  
%
Once a formal specification is written, it can be provided alongside the code as input to a verifier. A correct specification is essential, as it defines the behavior that the verifier attempts to prove. If the specification is incorrect or misaligned with the developer's intent, the verification process may succeed, but only for the wrong property. 
A good formal specification should be easy to write, review, and verify that it is valid on the given code.  

In our settings, formal specification of a method consists of a pre- and post-condition.
Within the context of the \texttt{MaxSub} problem, one natural formal specification is the post-condition of the \texttt{ensure} annotation of Listing~\ref{code:maxSubCodeImpl} (line 2), which uses the auxiliary definitions
in Listing~\ref{code:maxSubFormaSpec}. This method does not require a pre-condition.

\begin{lstlisting}[language=dafny, caption={Formal specification of the \texttt{maxSub} problem.}, label={code:maxSubFormaSpec},firstnumber=1,xleftmargin=15pt,numbersep=5pt]
function seqSum(ints: seq<int>): int {
 if |ints|====0 then 0 else ints[0] + seqSum(ints[1..])}

predicate IsMaxSubSum(ints: seq<int>, maxSum: int) {
 // a subarray exists with sum ==== 'maxSum'
 @exists s,e:: 0<=s<e<=|ints| && seqSum(ints[s..e]) ==== maxSum &&
 // all subarrays have sum <= 'maxSum' 
 forall s,e:: 0<=s<e<=|ints| ==> seqSum(ints[s..e]) <= maxSum}
\end{lstlisting}
The \texttt{seqSum} function (line~1) simply computes the sum of the elements of a given array in a recursive manner.\footnote{In Dafny, formal specifications are restricted to first-order logic and do not support imperative constructs such as \texttt{for} loops for value accumulation. State can only be propagated through the return values of recursive functions.} 
The \texttt{IsMaxSubSum} predicate (line~4) gets as input a sequence of integers and a \texttt{maxSum} value, checking \textit{(i)} that there exists a subsequence of \texttt{ints} starting at \texttt{s} and ending at \texttt{e} whose sum is exactly \texttt{maxSum} (line~6) and \textit{(ii)} that, for all possible subsequences of \texttt{ints}, the sum of the elements in each subsequence is no larger than \texttt{maxSum} (line~8).

\vspace{.05in}
Verifying that \texttt{MaxSubImpl} guarantees the \texttt{IsMaxSubSum} specification is non-trivial. The Dafny verifier cannot independently establish correctness, as formal verification often involves reasoning over an exponential number of implicit logical steps, a fundamentally complex task. To make verification feasible, users must assist the verifier by providing intermediate assertions, loop invariants, and supporting lemmas, which guide the verifier in generating more manageable subgoals within the verification process. We now discuss multiple challenges in generating such verification proofs and relate these challenges to AI-based systems. 

\subsection{Formal specification challenges}\label{ss:challenge-fspec}
\begin{figure}[t]
    \centering
    \includegraphics[width=.85\linewidth]{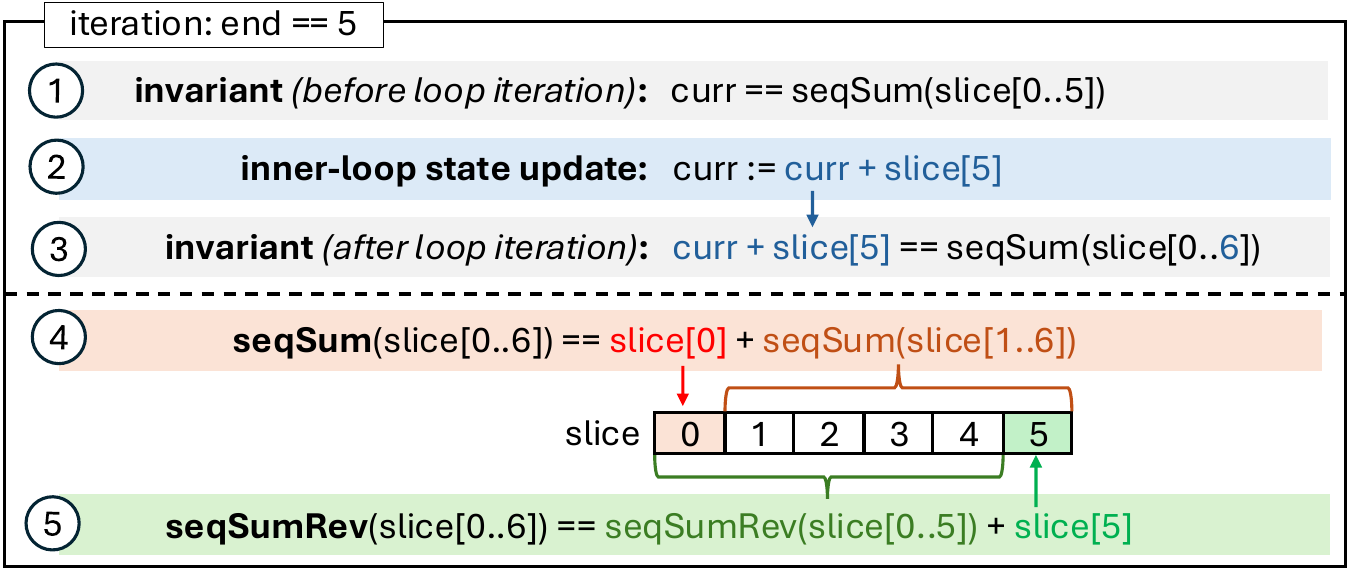}
    \vspace{-.13in}
    \caption{Inner-loop iteration with \texttt{end == 5}.}
    \label{fig:seqsumrev}
    \vspace{-.15in}
\end{figure}

\smartparagraph{Intuitive specifications may hinder verification.}
The way a specification is written can significantly impact the difficulty of the verification task: some formulations are more verifier-friendly than others.
Dafny fails, for example, to verify that  at the end of the inner loop, the value of \texttt{curr} is equal to the sum of the elements of the \texttt{slice} sequence, despite \texttt{curr} being increased exactly by each iterated element of \texttt{slice}. 
A user must provide a loop invariant and, with the support of Dafny, prove that it holds before entering the loop and is preserved by each iteration of the loop body. Once this is established, Dafny can verify the remaining proof obligations by assuming that the invariant holds at the start of every loop iteration and after the loop terminates.
In this case, the invariant should guarantee that the value of \texttt{curr} is equal to the sum of the elements analyzed so far:
\begin{lstlisting}[language=dafny,  numbers=none
, label={code:maxSubCode1}, firstnumber=1, xleftmargin=5pt, framexleftmargin=0pt,]
for end := 0 to |slice| 
 invariant curr == seqSum(slice[..end]) {  // invariant
  curr := curr + slice[end];
  maxSum := if curr > maxSum then curr else maxSum;
}   
\end{lstlisting}
To establish that the loop body preserves this invariant, we must show that, if the invariant holds \textit{before} executing the loop, then the invariant also holds \textit{after} the updated state of the program.
See an example in Fig.~\ref{fig:seqsumrev}, illustrated by points \circled{1}, \circled{2}, and \circled{3}.


Unfortunately, even after explicitly specifying this invariant, the latest version of Dafny (v4.10) is still not able to prove that, at the end of the loop, the current sum \texttt{curr} contains the sum of all elements of \texttt{slice}. The main reason lies in the way the formal specification is written, which we discuss next.

\smartparagraph{Misalignment between formal spec and code.} When taking a closer look at the \texttt{seqSum} specification, one can see that the calculation of the sum is recursive and starts from the end of the sequence (\ie the rightmost index) with the first index being the last element to be added. See point \circled{4} in Fig.~\ref{fig:seqsumrev}, where the first element is added as the last term in the summation. Unfortunately, the inner loop of \texttt{MaxSubImpl} computes the sum from the leftmost index. While it is obvious to a human that the sum will be identical, Dafny cannot know because of the  recursive structure of \texttt{seqSum}. 

\vspace{.05in}
One common solution involves rewriting the formal spec in the \textit{reverse} order so that the sum is computed from the leftmost element:

\begin{lstlisting}[language=dafny, numbers=none,
xleftmargin=5pt, framexleftmargin=0pt, 
% caption={Formal specification aligned with the code.}, 
label={code:maxSubFormaSpec-aligned}, firstnumber=1]
function seqSumRev(ints: seq<int>): int {
  if |ints| ==== 0 then 0 
  else ints[|ints|-1] + seqSumRev(ints[..|ints|-1])}
\end{lstlisting}
This approach aligns the verification process with the code (see point \circled{5} in Fig.~\ref{fig:seqsumrev}). 
In fact, in point \circled{3}, we replace \texttt{curr} with \texttt{seqSumRev(slice[..5])} that is equal to \texttt{seqSumRev(slice[..6])} based on point \circled{5}, proving the invariant for the next iteration.
However, this new formal specification also makes the specification less human readable and more prone to errors w.r.t. the original one.  

To avoid modifying the original specification, one can show that the two specifications are equivalent (i.e., proving \texttt{seqSum(ints) == \texttt{seqSumRev(ints)}} for all possible sequences \texttt{ints}). Once this equivalence is proved, a user can use the two specifications interchangeably. 
We argue that, as AI-based systems will keep improving, \textit{users should aim to write the simplest possible specification} that is both easy to verify and easy to review, even if this increases the burden of verifying the implementation against it, which future systems will mostly offload to AI-based automated systems. 

\begin{tcolorbox}[boxsep=0pt, boxrule=0pt, colframe=white, left=10pt, right=10pt, top=3pt, bottom=5pt]
\smartparagraph{Take-away 1. } Changing formal specifications to match the code can introduce errors. It is best to write clear, intuitive specifications, even if they misalign from the code, and rely on AI verifiers to bridge the gap.
\end{tcolorbox}

\subsection{Verifier-specific challenges}\label{sect:challenges-missing-assertions}
\smartparagraph{Quirks in formal verifiers can significantly slow down the verification process.}
In tools like Dafny, users are often required to prove small, seemingly trivial facts, such as sequence equivalence or set cardinalities, not because the logic is difficult, but due to limitations in how the underlying solver handles certain patterns. These quirks, rooted in the solver's internal heuristics, can make the verification process unexpectedly time-consuming.

For instance, consider again the proof of the invariant for the inner loop, this time using the updated formal specification \texttt{seqSumRev}. Even with this specification, Dafny still fails to verify the invariant autonomously. To assist the proof, the following assertion must be added to the loop body.
%
%

\begin{lstlisting}[language=dafny,  numbers=none
, label={code:maxSubCode2}, firstnumber=1, xleftmargin=5pt, framexleftmargin=0pt,]
for end := 0 to |slice|
 invariant curr ==== seqSumRev(slice[..end]) {   
  assert slice[..end+1][..|slice[..end+1]|-1] ==== slice[..end];
  curr := curr + slice[end];
  [...]
}   
\end{lstlisting}
In Dafny, assertions are used to guide the internal solver. They introduce a new proof goal that must be proven, but once verified, they can be used by the solver to prove the other proof goals. 

Even if the meaning of the assertion is trivial,
at first glance, it is not clear why this assertion is needed. To understand how this assertion relates to the verification of the invariant, one has to look at the logic deductions needed to prove the invariant. 



 \begin{lstlisting}[language=dafny, 
 % caption={Formal specification aligned with the code.}, 
 label={code:maxSubFormaSpec-aligned3}, numbers=none, xleftmargin=5pt, framexleftmargin=0pt, morekeywords={myBoldKeyword},]
seqSumRev(slice[..end+1]) 
  // by def of seqSumRev
  ==== if |slice[..end+1]| == 0 then 0 
     else slice[..end+1][|slice[..end+1]|-1] +
       seqSumRev(slice[..end+1][..|slice[..end+1]|-1]) 
  // by |slice[..end+1]| != 0
  ==== slice[..end+1][|slice[..end+1]|-1] +
     seqSumRev((*@\textbf{slice[..end+1][..|slice[..end+1]|-1]}@*))
  // by the assertion
  ==== slice[end] + seqSumRev((*@\textbf{slice[..end]}@*))
  // by the invariant
  ==== slice[end] + curr
\end{lstlisting}
%
We observed that, despite variations in code and formal specifications, Dafny often struggles to automatically verify this type of sequence equivalence. As a result, identifying and resolving these issues can be time-consuming and require manual reasoning with limited automation support.


There are even more unexpected cases that may require a disproportionate amount of time to fix. For instance, 
proving that $|A|<|B|$ when $A$ is a subset of $B$ appears to be trivial, however, a proof in Dafny requires to manually assert $|A-B| == 0$. These quirks derive from the fact that verifiers translate code to low level SMT formulas and feed these formulas into a solver, which is unaware of the original abstractions and semantics used by the verifiers.

\begin{tcolorbox}[boxsep=0pt, boxrule=0pt, colframe=white]
\smartparagraph{Take-away 2. }Many unexpected problems arise from quirks in verification tools rather than deep reasoning. Once these quirks are understood, solving tasks often becomes a matter of applying familiar time-consuming patterns, making it an ideal job for today's AI.
\end{tcolorbox}

\subsection{Challenges with proving lemmas}\label{ss:challenge-lemma}
Proving lemmas involves a range of reasoning effort, from straightforward facts that are obvious to humans to deeper logical insights that require strategic guidance. Our goal is to shift the burden of low-level, mechanically checkable reasoning to the AI verifier, allowing it to automatically verify lemmas that are obvious to humans. In contrast, complex reasoning steps, those requiring human insight or high-level understanding, should be expressed as proof outlines or hints provided by the user. This separation allows users to focus on expressing intent and structure, while the AI handles the tedious but routine logical steps.

As an example from the verification of the equivalence between \texttt{seqSum} and \texttt{seqSumRev}, a solution is to
show that \texttt{seqSum(ints)} can also be computed from the rightmost element, \ie \texttt{seqSum(ints) == seqSum(ints[..|ints|-1]) + ints[|ints|-1]} for any non empty sequences. This property cannot be proved automatically by Dafny.
The obvious way of proving such a lemma is by induction over sequence length. 
 \begin{lstlisting}[language=dafny, 
 % caption={Formal specification aligned with the code.}, 
 label={code:maxSubFormaSpec-aligned4}, numbers=none, xleftmargin=5pt, framexleftmargin=0pt, morekeywords={myBoldKeyword},]
lemma  lemmaSeqSumExtend(ints: seq<int>)
requires |ints| > 1
ensures seqSum(ints) ==== seqSum(ints[..|ints|-1]) + ints[|ints|-1]
{ if |ints| > 2 {
    lemmaSeqSumExtend(ints[1..]);
    assert ints[1..][..|ints|-2] ==== ints[1..|ints|-1]; }
}
\end{lstlisting}
These proofs for intuitive properties often rely on applying familiar inductive patterns with limited reasoning capabilities, a task amenable to AI. Note the extra assertion 
for sequence equivalence.


\begin{tcolorbox}[boxsep=0pt, boxrule=0pt, colframe=white]
\smartparagraph{Take-away 3.} Lemmas that require limited reasoning and are obvious to humans can often be verified automatically by AI, while more complex proofs benefit from human-provided outlines that guide the AI through the harder steps.
\end{tcolorbox}

\subsection{Challenges with complex code}\label{ss:challenge-complex}
Today's AI-based verifiers can verify small, self-contained fragments of code, particularly when the required reasoning is low and it aligns closely with the structure of the code. However, their reasoning capabilities remain limited and fail when the code grows in size or structural complexity. Taking the two nested loops of \texttt{MaxSubImpl} in Listing~\ref{code:maxSubCodeImpl} as an example, to completely verify the implementation, we must add three invariants to the nested loop: \first there is indeed a subsequence whose sum is \texttt{maxSum} that we computed until now; \second all subsequences starting before the index that is currently processed by the outer loop (\ie \texttt{start}) have sum not greater than \texttt{maxSum}; and \third all subsequences starting from the current index of the outer loop and ending before the current index of the inner loop (\ie \texttt{end}) have sum not greater than \texttt{maxSum}.

\begin{lstlisting}[language=dafny, 
, label={code:maxSubCode3}, firstnumber=1,xleftmargin=15pt,numbersep=5pt]
for start := 0 to |ints|
[...]
  for end := 0 to |slice|
    invariant @exists s,re :: 0 <= s <= e <= |ints| && 
              seqSum(ints[s..e]) == maxSum
    // previous outer loop iterations  
    invariant forall s,e :: 0 <= s < e && r <= |ints| ==> 
              seqSum(ints[s..e]) <= maxSum
    // current outer loop iteration
    invariant forall s,e :: s == start && e <= s + end ==> 
              seqSum(ints[s..e]) <= maxSum
{ ... }   
\end{lstlisting}
We note that the invariants in the inner loop must be specified with respect to \texttt{ints} instead of the more natural \texttt{slice}.
Coupling the invariants of the inner loop with the logic of the outer loop, even in this simple example, becomes overwhelming.
In fact, 
an AI-based system should overcome multiple difficulties simultaneously. 
\begin{itemize}[leftmargin=*,noitemsep]
    \item Derive the missing subsequence equivalence assertions. 
    \item Prove the mapping of elements between \texttt{ints} and \texttt{slice}.
    \item Derive the correct invariants for the inner and outer loops.
    \item Potentially define a simpler formal specification and prove its equivalence to the original one. 
\end{itemize}

\noindent While an AI-based system may be able to complete the verification of a single task above, it becomes much more challenging when all of them are involved. In fact, at the beginning of the verification process, the feedback from the verifier may not be particularly useful (\ie it limits to inform that the method post-condition cannot be proved) and an AI may easily be derailed on the wrong path. Based on our experiments, cutting-edge LLMs are unable to verify \texttt{MaxSubImpl} using the iterative feedback from the Dafny verifier. They consistently fail to handle the complexity of the code, particularly when it extends beyond a simple \texttt{for} loop iteration.


\begin{tcolorbox}[boxsep=0pt, boxrule=0pt, colframe=white]
\smartparagraph{Take-away 4.}     
     AI-based systems struggle to verify complex code due to the vast search space and limited feedback from the verifier when key proof elements are missing. Working with smaller, well-scoped code fragments reduces this complexity, enabling AI to make meaningful progress in a step-by-step manner.
\end{tcolorbox}

\section{\systemname: Transiently Simplifying Code}\label{sec:system}

\begin{figure*}[t]
    \centering
    \includegraphics[width=\linewidth, trim={0 .4in 0 0}]{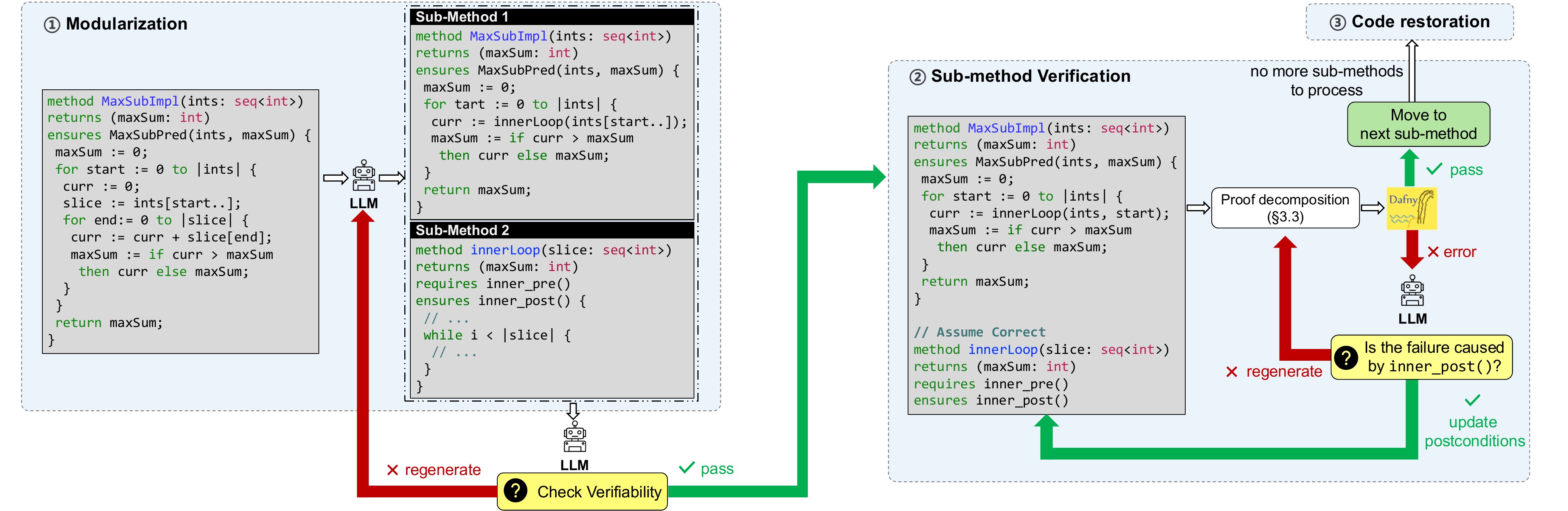}
    \caption{Example of code-level decomposition.}
    \vspace{-.13in}
    \label{fig:prometheus_loop_lift}
\end{figure*}

We argue that LLMs remain a natural candidate to automate and facilitate code verification, but they require \textit{guidance} when handling non-trivial algorithms to address the gaps mentioned in Sec.~\ref{sec:motivation}. 

To address the limitations of LLMs, such as hallucinations and limited reasoning, we introduce \systemname, an LLM-assisted system that uses a decomposition-based strategy operating along two complementary dimensions: code and proof. Our central insight is to shift the focus from verifying the original code at all costs to allowing the AI to first obtain \textit{some verified version} of it, even if that means modifying the code. The first contribution lies in transforming the code into a variant that is easier and faster for the LLM to verify, steering it away from unprovable or overly complex paths. The second contribution focuses on the proof itself: it decomposes complex lemmas into smaller, more manageable ones that are easier for the LLM to infer and verify. Once the adjusted code is verified, it serves as a solid foundation and reliable ground truth from which the proof can be incrementally adapted back to the original code. Additionally, to support deeper reasoning and reduce hallucinations, we optionally allow users to provide natural language proof sketches that outline the intended verification strategy.




The code decomposition (\S\ref{ss:code-decomposition}) addresses the challenges described in Sec.~\ref{ss:challenge-complex} by increasing the focus of the AI on smaller, more manageable parts. We perform \textit{loop lifting} by refactoring the imperative loops into named functions that encapsulate their bodies. This enables \systemname to isolate the control flow and eliminate interactions between loops that could hinder verification tasks due to intermediate invariants between multiple nested loops.

The proof decomposition (\S\ref{ss:proof-decomposition}) tackles the challenges outlined in Sec.~\ref{ss:challenge-lemma}. \systemname breaks down complex proof obligations into a hierarchy\slash tree of structured helper lemmas and sub-lemmas. These lemmas capture inductive steps or auxiliary properties that are generally difficult for LLMs to infer. The goal of this decomposition is to enable LLMs to focus on simpler, localized proof sub-goals, aligning with the insights of Sec.~\ref{sect:challenges-missing-assertions}. 
Proof decomposition also plays a key role in addressing challenges introduced by formal specification misalignments (Sec.~\ref{ss:challenge-fspec}). By isolating verification into smaller proof goals, \systemname makes it possible to bridge these gaps incrementally, without requiring the specification to be rewritten.

\vspace{-.05in}
\subsection{Challenges}
Realizing this vision of modular, decomposition-driven verification is non-trivial. While code- and proof-level decomposition offer a structured pathway toward scalable reasoning, putting them into practice introduces several technical challenges. 

\smartparagraph{Decomposition versus restoring complexity.} While decomposing code into smaller units can simplify verification, it can obscure the connection to the original structure, making it difficult to reconstruct a coherent proof for the original code. This tension is more pronounced when transformations introduce auxiliary variables, reorder logic, or isolate control flow. \systemname explores various decomposition strategies along with techniques to 
restore the original code and transpose the verified proofs back onto it.




\smartparagraph{Pruning unverifiable proof paths.} Proof decomposition can dramatically expand the search space, and deciding \emph{when}, \emph{where} to split obligations is a hard problem. Existing LLM-driven proof synthesis operates under uncertainty and frequently proposes speculative sub-lemmas. However, naive decomposition heuristics can mislead the verification process, suggesting lemmas that are not verifiable. 
A key challenge for \systemname is to use verifier feedback, heuristics, and LLMs to identify wrong branches early and cut them off before they waste significant resources.

\smartparagraph{Verifiability versus usefulness.} Automatically synthesized helper lemmas may be 
irrelevant, overly strong, or based on overly restrictive assumptions, or not useful. 
In other words, a successfully verified lemma might not help prove the original formal specification. \systemname must assess not only verifiability but \emph{utility} as well. Determining whether a lemma contributes meaningfully to the final goal is a core challenge of \systemname.


\smartparagraph{Recovering from failures and reusing partial progress.} Checking the verifiability of each module is also hard, even with strong guidance, AI-based tools cannot guarantee that proof search always follows a verifiable decomposition path. \systemname may commit to an unproductive branch and fail. However, portions of the generated proof developed along the exploration may still be correct and reusable. 
Therefore, how to detect the failure early and selectively roll back or redirect the search while preserving validated lemmas and proof fragments is a key challenge.





\subsection{Code-level decomposition}\label{ss:code-decomposition}
\smartparagraph{\circled{1} Modularization.} As discussed in Sec.~\ref{ss:challenge-complex}, the length and the nesting depth of large programs can significantly hinder verification. While Dafny can easily generate proofs in sequential code, reasoning with loops necessitates specifying invariants. This is because the verifier cannot deduce the number of loop iterations or the properties maintained within the loop body. Consequently, omitting even a single invariant  leads to verification failures and makes it difficult to detect the source of the error.

%
To tackle this problem, \systemname prompts an LLM to break the code at loop-level modularity, \ie  to extract auxiliary sub-methods, each containing at most one loop, and rewrite the original method to call them. Fig.~\ref{fig:prometheus_loop_lift} shows an example where \systemname transforms the nested loop \texttt{MaxSubImpl} into two methods: \first \texttt{MaxSubImpl}, which keeps the original signature and contract, and \second \texttt{innerLoop}, equipped with  newly‑generated signatures \texttt{innerPre} and \texttt{innerPost}. 
%
We perform a sanity check by asking an LLM to ensure consistency \ie that both generated methods are verifiable and their contracts suffice to restore the original code. These checks allow \systemname to filter out most unverifiable cases before entering an unverifiable path. \systemname refines the decomposition until succeeds or reaches a configurable attempt limit.

\smartparagraph{\circled{2} Sub-method verification.} Each sub-method is verified sequentially by the proof-level decomposition described in Sec.~\ref{ss:proof-decomposition}. 
When Dafny verification fails, the cause often falls into one of two categories: \first an insufficient postcondition in a called method \eg the postcondition of \texttt{innerPost} may be too weak to support the verification of \texttt{MaxSubImpl} (\eg \texttt{ensures true}), or \second missing intermediate proof steps within the current method. In the former case, \systemname prompts the LLM to strengthen the callee's contract; in the latter, the system proceeds with further proof-level decomposition within the current method.

\smartparagraph{\circled{3} Code restoration.} Once all sub-methods are verified, \systemname reconstructs the original code by \first using an LLM to merge the  sub-methods into a single cohesive method, and \second mapping the verification code back to the original structure. During this process, minor verification inconsistencies are automatically addressed by incorporating Dafny verifier's feedback into subsequent LLM prompts. The final output is a program that is both functionally equivalent to the original input and fully verified. If \systemname fails to restore the original code automatically, the user can manually map the verified methods back to the original code, which still remains significantly easier than verifying the code from scratch.

\begin{figure*}[t]
    \centering
    \includegraphics[width=\linewidth, trim={0 .5in 0 0}]{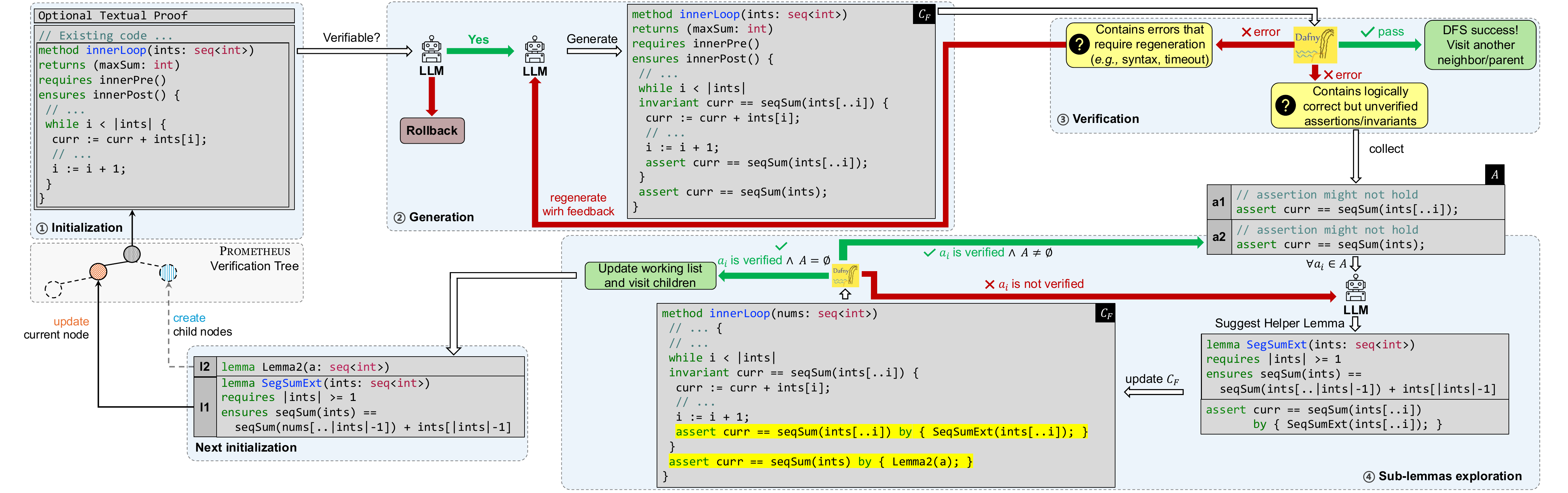}
    \caption{Example of proof-level decomposition.}
    \label{fig:prometheus}
    \vspace{-.15in}
\end{figure*}

\vspace{-.1in}
\subsection{Proof-level decomposition}\label{ss:proof-decomposition}

Fig.~\ref{fig:prometheus} shows an example of proof-level decomposition performed by \systemname on the \texttt{innerLoop} method.

\smartparagraph{\circled{1} Initialization Step.} Starting from the input code, \systemname begins analyzing and generating the missing annotations (\ie assertions, lemmas, invariants) to successfully verify the provided code.
In particular, \systemname starts building a tree data structure that serves to store and evaluate each step of the verification process. 
We use the input information to create the root node, initiating the automatic generation process.
Each tree node corresponds to a sub-lemma that \systemname must verify to validate the correctness of the main method. A tree node contains \first a partially verified version of the code, \second the signature of the sub-lemma including pre- and post-conditions, and \third the optional textual proof.


After creating the root node, we begin the tree traversal from it. Note that once the traversal is initiated, \systemname operates automatically \textit{without requiring} any further human intervention.


\smartparagraph{\circled{2} Generation Step.} When visiting a node, \systemname first creates a prompt for the LLM using the proof text (if provided) and the existing code, 
asking it to evaluate at a high level whether the signature is logically correct\slash verifiable. Specifically, the model is requested to provide a Boolean response (yes\slash no). 
If the signature is verifiable, \systemname proceeds to construct a more detailed prompt instructing the LLM to generate all the necessary fragments for verification. The required fragments depend on the specific input provided to the LLM. If only the signature is supplied as input, \systemname requests both the code body and all required verification fragments. Conversely, if a code body is already provided, \systemname prompts the LLM to enhance the existing code by adding any missing verification fragments. 

Successful verification produces a code snippet for the given signature (denoted $\overline{\mathcal{C}}$), which is merged into the existing tree-node code to form a verified version $\mathcal{C_F}$. \systemname then initiates a verification phase to confirm correctness.

\smartparagraph{\circled{3} Verification Step.} \systemname runs the Dafny verifier with $\mathcal{C_F}$ as input and analyzes the output. Using the verifier's feedback, the system identifies the errors present in the generated code and takes steps to address them. For certain error types (\eg syntax errors, unprovable pre- or post-conditions, or verifier timeouts), the system does not attempt direct fixes. Instead, it re-prompts the LLM to produce a revised version of $\mathcal{\overline{C}}$. During this process, it incorporates feedback derived from the verifier's output and includes additional prompt guidance to assist the model in identifying and resolving the issue, reducing the likelihood of generating the same version of the code.
These additional prompts, carefully crafted by us based on common general Dafny issues and potential resolution strategies, are never specific to any instance in the dataset.
%

\systemname focuses on automatically repairing two types of errors: \first assertions that are presumably valid but cannot be proven by Dafny, and \second invariants that are presumably correct but cannot be verified within loop bodies. We focus on assertions and invariants since, beyond the pre- and post-conditions, they are the most important annotations within the context of a method or a lemma.

Beginning with assertions, the system identifies all assertions in $\mathcal{\overline{C}}$ that fail to hold. For each assertion, \systemname prompts the LLM to evaluate whether the assertion is logically correct and if an additional sub-lemma could assist in proving the assertion. If so, the system requests the LLM to generate the sub-lemma's signature and its corresponding call (\ie the sub-lemma with appropriate arguments). Upon successful generation, starting from $\mathcal{C_F}$, the sub-lemma call is added to the failing assertion line using the \texttt{by \{ ... \}} notation, and the lemma signature is appended to $\mathcal{C_F}$. The Dafny verifier is then invoked again, and any resulting errors initiate another attempt to generate assertions. This process is repeated a maximum of $t$ times, where $t$ is user-configurable. Successful verification of $\mathcal{C_F}$ indicates that a new sub-lemma must be verified. After all the errors are resolved, \systemname maintains, for each tree node, a working list that tracks the signatures of all newly generated sub-lemmas. The above process stops when all the assertions have been processed.

For invariants, \systemname follows a similar approach. It identifies all invariants in $\mathcal{\overline{C}}$. \systemname leverages the LLM to determine whether each failed invariant is logically verifiable. If so, \systemname prompts the LLM to generate an additional sub-lemma to prove it. The sub-lemma must contain a lemma signature that asserts the
invariant holds after one iteration of the loop. Upon successful generation, the sub-lemma call is automatically inserted into the code, and the lemma signature is appended to $\mathcal{C_F}$. Also in this case, \systemname allows up to $t$ attempts to generate an appropriate sub-lemma before adding it to the node working list. The process terminates once all the invariants have been processed.

\smartparagraph{\circled{4} Sub-Lemmas Exploration.} Successfully visiting a tree node results in a verified code body $\mathcal{\overline{C}}$ for the node's signature, along with one or more new sub-lemma signatures that require further verification (tracked in the working list). The signatures and calls to these sub-lemmas are already correctly placed and verified in the resulting code. This intermediate code $\mathcal{C_F}$ represents an updated version that is verified but lacks the body of those sub-lemmas that must still be visited. Therefore, the process must be repeated for all sub-lemmas in working list, until no additional code remains to be verified. It is important to note that during the verification of a new sub-lemma, further sub-lemmas may still be generated and required, making the process recursive. 

At this stage, \systemname initiates a recursive generation and exploration of the children of the current visited node. For each sub-lemma signature $l_k$ in the working list, a corresponding tree node is created and initialized as follows: the parent node is set to the current node, the starting code is set to the current working code $\mathcal{C_F}$, the sub-lemma signature is $l_k$, and the textual proof is empty. A Depth-First Search (DFS) is started on the first child node.

\smartparagraph{Rollback.} 
Since \systemname runs fully autonomously, it may need to discard an entire tree when some lemma signatures are unverifiable. This occurs when the LLM fails to generate a verifiable code block after $t$ attempts (user-defined). To avoid losing all progress, \systemname includes a fallback: on repeated generation or verification failure, a node can be retried up to $s$ times (also user-defined), and each retry lowers the LLM’s temperature by $\Delta \mathit{temp}$.
If all the $s$ attempts fail, the current node's verification in the DFS is terminated, and the system rolls back to its parent, which also has $s$ chances to generate a proof. This process continues iteratively up to the root. If the root also fails after $s$ attempts, the entire generation process is aborted. Briefly, the approach resembles simulated annealing~\cite{simulated-annealing}: we begin with a high generation temperature to encourage creative solution from the LLM, and gradually lower it if failures persist, guiding the model toward more deterministic outputs. The process ends when the temperature reaches~$0$, ensuring fully deterministic generation.
\section{Evaluation}\label{sec:evaluation}

In this section, we present a comprehensive evaluation to showcase the performance of \systemname. All code, datasets, and prompts can be found in the online artifact. We conducted several experiments designed to answer the following research questions:

    \vspace{.05in}
    \noindent \textbf{Q1: How does \systemname perform?} We evaluate \systemname' capabilities to automatically generate proper invariants, assertions, and helper lemmas to verify a program.
    
    \noindent \textbf{Q2: Is \systemname robust to different code decomposition strategies?} We aim to evaluate the influence of different code decomposition strategies on \systemname.
    
    \noindent \textbf{Q3: Is \systemname robust across formal specifications?} We assess how \systemname handles challenges posed by different formal specifications for the same problem, as described in Sec.~\ref{sec:motivation}.
    
    \noindent \textbf{Q4: Does \systemname scale to verify non-trivial programs?}
    We examine \systemname performance as the size and complexity of program verification tasks grow. 
    


\smartparagraph{Selected datasets.} Due to the limited exploration of LLM-assisted Dafny solutions, there are only three existing datasets available.
%
MBPP-DFY-153~\cite{misu2024towards} is a dataset composed of $153$ Dafny programs, most of which consist of a single Dafny method. 
To align with our goal of evaluating LLMs' performance in generating verification code, we made these modifications:
\first~we removed all the verification-related code except pre- and post-conditions (\eg invariants, assertions, lemmas), and \second~we ran the Dafny verifier and retained only the programs whose verification failed (\ie programs that Dafny cannot verify without additional verification code). After this process, we obtained $90$ unverified programs, which is our dataset for the evaluation.
DafnyGym~\cite{mugnier2024laurel} is a dataset comprising lemmas extracted from real-world Dafny codebases. We chose not to include it in our evaluation, as the lemmas are relatively simple and comparable to MBPP-DFY-153, \eg involving only a single missing assertion line.
DafnyBench~\cite{loughridge2024dafnybench} is a benchmark comprising all existing Dafny code available on GitHub. We do not evaluate using DafnyBench since the quality of the benchmark is notably inconsistent. This stems from the nature of the dataset: \first~code written for different domains without clear problem statements (\eg lemma libraries, tutorial references, or industrial programs designed to verify complex algorithms), and, most importantly, \second~code that fails even to pass Dafny's syntax checks. 

To assess \systemname's ability to handle \textit{more complex} programs, we designed a new Dafny benchmark. We select multiple programming tasks involving arrays from LeetCode~\cite{leetcode} and ask GPT-4 to generate a naive (often brute-force) implementation in Dafny using test cases. We obtain $36$ programs and we ask GPT-4 to generate a formal specification for each of them. It is not trivial to automatically test a formal specification as it requires a proof of correctness as input. Thus, we manually verify the correctness of the formal specification of $22$ programs ($12$ including at least one nested loop), and call the dataset \emph{TitanBench}. 
\footnote{Manual verification is time consuming; verifying a single instance can take several hours. Therefore, we leave the expansion of TitanBench as future work.}

\smartparagraph{Baseline: LLM with detailed feedback.} We evaluate \systemname by comparing it with a baseline approach that iteratively feeds an LLM with feedback from the Dafny verifier, as proposed in recent work on code verification~\cite{sun2024clover,mirchev2024assured,parnian2024pccc}. For fairness, we use the same LLM model that we use in \systemname.
As a production-grade tool, Dafny often returns specific error reports (\eg ``line X: the invariant cannot be proved''). When the LLM produces an incorrect proof, we append the feedback from the Dafny verifier to the original conversation with the LLM and ask it to fix the issue based on the error message. 
We do not consider complementary techniques such as the dynamic few-shot approach proposed by Misu \textit{et al.}~\cite{misu2024towards}, which rely on retrieving similar code examples from a database and include them as few-shot prompts.
In this paper, we focus on the reasoning ability of LLM-based systems to verify code, possibly guided by a proof outline, but without relying on retrieval of code snippets, which is an orthogonal problem.
We also tested VerMCTS~\cite{brandfonbrener2024vermcts}, an early-stage tool using an advanced MCTS-based approach for Dafny. Unlike our work, which focuses on verification, VerMCTS jointly generates specifications, code, and proofs, but consistently failed to produce correct results, especially specifications, with Claude Sonnet 3.7.

\smartparagraph{LLMs selection.} We primarily use Claude Sonnet 3.7~\cite{sonnet3.7}, 
as the Claude series models have demonstrated superior performance in generating Dafny code~\cite{loughridge2024dafnybench}. Claude Sonnet 3.7 was the latest available version when this work began, ensuring it was not exposed to our experiments or prompts. Additionally, we utilize OpenAI o4-mini~\cite{o4} 
for checking code verifiability, as it excels in complex reasoning at the time of submission.
%
All the models run with the following configuration: initial temperature of $0.5$, the maximum number of tokens is $4028$, and the timeout is $20$\,s. We selected these values to strike a balance between creativity, concise responses, and reasonable generation time.

\smartparagraph{Evaluation configuration.} To prevent infinite verification loops, we set Dafny's verification timeout to $20$ seconds. Within \systemname, each node of the tree is allowed a maximum of $t=10$ generation attempts before rolling back. The regeneration counter $s$ is set to $2$, and $\Delta {temp}$ to $0.3$. This means that each node has two generation attempts with temperatures of $0.5$ and $0.2$. Finally, we set a hard timeout of $500$ seconds on the generation process for both the baseline and \systemname, after which the attempt is aborted.



\smartparagraph{Evaluation metrics.} We evaluate the performance of LLMs in code verification using the success rate, defined as the percentage of successful runs out of all attempts. In particular, we report the success rate with the \textit{verify@5} metric, allowing LLMs up to 5 attempts to generate the correct code. 
In the case of \systemname, a single execution of \systemname on a program is considered as one run. The trends in our results hold even by increasing the number of attempts due to the inherent complexity of the verification tasks.

\smartparagraph{Influence of training dataset on the results.} To the best of our knowledge, no equivalent dataset to our newly introduced TitanBench exists, meaning that models like Claude Sonnet or OpenAI’s GPT have not been trained on it, ensuring unbiased results. Even if we were to assume that some LLMs have been trained on these benchmarks (or equivalent ones), we still make a crucial observation: \textit{the baseline fails to solve any algorithm that involves more than one of the challenges outlined in Sec.~\ref{sec:motivation}}.


\subsection{Verification Performance (Q1)}\label{ss:rq1}

\begin{table}[t]
\centering
\caption{Comparison over TitanBench and MBPP-DFY-153.}
\vspace{-.13in}
\resizebox{\linewidth}{!}{%
\begin{tabular}{c c c c c c c c c}
 & \multicolumn{6}{c|}{\textbf{TitanBench}} &  \\
 \cline{2-7}
 & \multicolumn{2}{c|}{\makecell{Overall \\ (22)}} &\multicolumn{2}{c|}{\makecell{w/ nested loop \\ (10)}} & \multicolumn{2}{c|}{\makecell{w/o nested loop \\ (12)}} &  \multicolumn{2}{c}{\multirow{-2}{*}{\makecell{\textbf{MBPP-DFY-153} \\ (90)}}} \\
 & \# & \multicolumn{1}{c|}{Succ. Rate} & \# & \multicolumn{1}{c|}{Succ. Rate} & \# & \multicolumn{1}{c|}{Succ. Rate} & \# & Succ. Rate \\
 \cline{1-9}
 Baseline &
 15 & \multicolumn{1}{c|}{68\%} &
 6 & \multicolumn{1}{c|}{60\%} &
 \textbf{9} & \multicolumn{1}{c|}{\textbf{75\%}} &
 89 & 98\% \\
 \cline{1-9}
 \systemname &
 \textbf{19} & \multicolumn{1}{c|}{\textbf{86\%}} &
 \textbf{10} & \multicolumn{1}{c|}{\textbf{100\%}} &
 \textbf{9} & \multicolumn{1}{c|}{\textbf{75\%}} &  
 \textbf{90} & \textbf{100\%} \\
\end{tabular}
}
\label{tab:rq1}
\vspace{-.15in}
\end{table}

To answer the first research question, we evaluate \systemname against the selected baseline using the $22$ tasks from TitanBench and the $90$ selected programs from MBPP-DFY-153. 





\smartparagraph{\systemname achieves higher success rate on tasks involving nested loops and complex reasoning.} We begin by demonstrating \systemname's ability to tackle non-trivial verification tasks using the TitanBench benchmark. As shown in Table~\ref{tab:rq1}, our system successfully completes $19$ out of $22$ tasks, achieving a success rate of 86\%, compared to the baseline's 68\%. Among the $22$ tasks, $12$ involve programs with at least one nested loop. Importantly, the four additional successful verifications by \systemname come from these nested loop tasks. While the baseline solves $6$ out of $10$ such tasks, \systemname successfully \textit{solves all of them}. Upon reviewing the solutions provided by \systemname for these four tasks, we found that verifying them requires either \first multiple invariants for each loop, or \second additional helper lemmas beyond a single method. 

\smartparagraph{Providing feedback on errors enhances the baseline success rate, yet decomposition is key to full verification.} We now perform a comparison on the MBPP-DFY-153 dataset. Since the dataset primarily consists of simple single-method Dafny programs, with the verification code generally under four lines, the baseline approach already achieves an extremely high success rate, solving $89$ out of $90$ tasks, as shown in Table~\ref{tab:rq1}. After analyzing the LLM generation logs, we noticed that the feedback from the Dafny verifier helps the LLM correct its errors. For example, with a syntax error, the LLM generates a revised response with corrected syntax. 
We also observe that the baseline occasionally attempts to solve a verification problem by proposing helper lemmas. 

However, even when it successfully suggests such lemmas, the baseline is likely to fail because it must address both the original method verification and the new lemma verification tasks simultaneously. In the one failed task, the baseline successfully proposes the key lemma required for proving the main method but fails to verify it after several attempts. Conversely, \systemname solves the task by breaking it into smaller, more manageable sub-lemmas and verifying them individually. As shown in Table~\ref{tab:rq1}, \systemname successfully verifies \textit{all tasks} in MBPP-DFY-153. 

\vspace{-0.05in}
\subsection{Robustness to decomposition strategies (Q2)}\label{ss:rq2}

As described in Sec.~\ref{sec:motivation}, \systemname tackles complex programs, such as those involving nested loops, by decomposing the code into separate components that are easier to verify. The decomposition strategy may influence \systemname' verification performance. To explore this, we select a subset of tasks from TitanBench involving nested loops that can be successfully verified by \systemname. For each task, we generate three variants based on three distinct decomposition strategies: \first \textbf{Full-Sharing}: Intermediate results from the outer loop are passed to the inner loop; \second \textbf{Decoupled}: Intermediate results are not passed, but all input variables in the outer loop are passed; \third \textbf{Fully-Decoupled}: Intermediate results are not passed, and only the relevant input variables are passed. 
Taking Listing~\ref{code:maxSubCodeImpl} as an example, \textit{Full-Sharing} passes the entire state to the inner loop, including the current \texttt{maxSum}, \textit{Decoupled} passes only necessary variables \texttt{ints} and the index \texttt{start}, while \textit{Fully-Decoupled} only passes the sliced sequence \texttt{ints[start..]}.

This process results in a collection of eight examples, each having three variants. All variants preserve the original specification, with only the implementation differing.
Since these tasks involve verifying more than one method and \systemname typically begins the code generation from the root of the tree, we initialized the tree before starting the DFS as follows: \first we ask the LLM to provide an initial signature for the inner method; \second we reconstruct the tree by adding the outer method as the root node and the inner method as a child node in its working list; and \third we run \systemname based on this reconstructed tree. This initialization ensures that while verifying the outer method, \systemname can still update the signature and check the verifiability of the inner method.

Table~\ref{tab:rq2} shows the success rates for each decomposition strategy. 
In all cases, \systemname demonstrates robust performance, verifying both methods with a high success rate.  We note that no single approach succeeds in all cases. Since all tasks were drawn from successful runs, \systemname must rely on different strategies depending on the task.
Interestingly, increasing the modularity does not necessarily result in a simplified verification process. With the \textit{Fully-Decoupled} decomposition approach, the verification process becomes more challenging despite the better isolation of the inner loop.  Manual inspection reveals that it is often difficult to relate the property verified on the sub-array back to the original array in the outer loop, resulting in significantly more effort compared to the other two strategies.

We also tested our restore mechanism on the successful runs. The results demonstrate that all verified methods can be successfully merged and reconstructed into the original code structure.

\begin{table}[t]
\centering
\caption{Success rates of verification\slash restoration with different decomposition strategies.}
\vspace{-.13in}
\resizebox{.8\linewidth}{!}{%
\begin{tabular}{c c c}
    \textbf{Decomposition Strategy} &
    \textbf{Verification} &
    \textbf{Restoration} \\ \hline
    Full-Sharing    & 87.5\% (7/8) & 100\% (7/7) \\
    Decoupled    & 87.5\% (7/8) & 100\% (7/7) \\
    Fully-Decoupled & 62.5\% (5/8) & 100\% (5/5)
\end{tabular}
}
\label{tab:rq2}
\vspace{-.2in}
\end{table}
\subsection{Robustness across formal specifications (Q3)}\label{ss:rq3}

As discussed in Sec.~\ref{ss:challenge-fspec}, changing formal specifications to match the code can introduce subtle errors that are difficult for humans to detect. \systemname addresses this challenge by generating helper lemmas that bridge the misalignment between the formal specification and the code. 
To evaluate this capability, we select all $13$ tasks successfully verified by the baseline in TitanBench and modify their specifications in three ways: \first by introducing or reversing structural recursion; \second by incorporating more detailed definitions involving complex set computations; and \third by intentionally introducing a mismatch between the predicates and the code. The implementations themselves are left unchanged. 
%

\smartparagraph{\systemname effectively handles specification-code misalignment, especially in the presence of nested loops.} Table~\ref{tab:rq3} shows the results of the experiment, with tasks categorized by the presence\slash absence of nested loops.
Among the 13 tasks, the baseline successfully verifies the 4 simplest ones. Notably, these tasks do not involve any arithmetic operations over the input array, which significantly simplifies verification. Without any arithmetic aggregation across the elements of the array, the inner loop does not need to maintain or reason about an evolving state, making the proof much more tractable.
However, the baseline struggles significantly with tasks that include nested loops and aggregate state, aligning with our observations in Sec.~\ref{ss:challenge-fspec} and Sec.~\ref{ss:challenge-complex} that AI-based systems tend to fail when multiple reasoning steps are required and to deal with misaligned, yet more readable, formal specifications.
Conversely, \systemname substantially outperforms the baseline by verifying $9$ among the $13$ tasks, including those with nested loops. \systemname' code decomposition mechanism enables it to modularize loops effectively, while its proof decomposition component introduces the necessary helper lemmas. Together, these features allow \systemname to verify $4$ tasks that involve nested loops. In the three failed cases, \systemname succeeds in proposing helper lemmas to bridge the misalignment, but fails in proving them. These lemmas are not trivial and require more advanced reasoning.  We leave addressing this challenge for future work.

\begin{table}[h]
\vspace{-.1in}
\caption{\systemname successfully deals with various specification definitions.}
\vspace{-.13in}
\resizebox{.9\linewidth}{!}{%
\begin{tabular}{c c c c}
 & \textbf{wo/ Nested Loop} & \textbf{w/ Nested Loop} & \textbf{All} \\ \hline
Baseline   & 42.8\% (3/7) & 14\% (1/7) & 30\% (4/13) \\
\systemname &  \textbf{71.4\% (5/7)} & \textbf{57\% (4/7)} & \textbf{69\% (9/13)}
\end{tabular}
}
\vspace{-.13in}
\label{tab:rq3}
\end{table}

\begin{table}[t]
\caption{Improved performance on non-trivial tasks.}
\vspace{-.13in}
\resizebox{\linewidth}{!}{%
\begin{tabular}{c c c c c c c c}
 & &    & \multicolumn{2}{c|}{Baseline}     & \multicolumn{3}{c}{\systemname}                             \\ \cline{4-8} 
\multirow{-2}{*}{\#} &
  \multirow{-2}{*}{LOVC} &
  \multirow{-2}{*}{\makecell{N. \\Lemmas}} &
  \multicolumn{1}{c}{Success} & \multicolumn{1}{c|}{\makecell{Avg. Time \\(s)}} &
  \multicolumn{1}{c}{Success} &
  \multicolumn{1}{c}{\makecell{Avg. Time \\(s)}} &  \makecell{Avg. \\N. Lemmas} \\ \hline
    1 & 21  & 1  & \multicolumn{1}{c}{\textbf{5/5}} & \multicolumn{1}{c|}{23.06}  & \multicolumn{1}{c}{\textbf{5/5}} & \multicolumn{1}{c}{73.20}  & 2.3 \\
2 & 69  & 1  & \multicolumn{1}{c}{1/5} & \multicolumn{1}{c|}{1191.80} & \multicolumn{1}{c}{\textbf{3/5}} & \multicolumn{1}{c}{409.30} & 6   \\
3 & 140 & 1  & \multicolumn{1}{c}{0/5} & \multicolumn{1}{c|}{-}      & \multicolumn{1}{c}{\textbf{3/5}} & \multicolumn{1}{c}{299.60} & 1   \\
4 & 171 & 3  & \multicolumn{1}{c}{0/5} & \multicolumn{1}{c|}{-}      & \multicolumn{1}{c}{\textbf{2/5}} & \multicolumn{1}{c}{725.50} & 12  \\
5 & 236 & 9  & \multicolumn{1}{c}{0/5} & \multicolumn{1}{c|}{-}      & \multicolumn{1}{c}{\textbf{3/5}} & \multicolumn{1}{c}{369.00}   & 8.3 \\
6 & 245 & 5  & \multicolumn{1}{c}{0/5} & \multicolumn{1}{c|}{-}      & \multicolumn{1}{c}{\textbf{1/5}} & \multicolumn{1}{c}{793.00}   & 18  \\
7 &  285 &  5 &  \multicolumn{1}{c}{0/5} & \multicolumn{1}{c|}{-} &  \multicolumn{1}{c}{\textbf{1/5}} &  \multicolumn{1}{c}{1396.00} &  14 \\
8 & 487 & 11 & \multicolumn{1}{c}{0/5} & \multicolumn{1}{c|}{-}      & \multicolumn{1}{c}{0/5} & \multicolumn{1}{c}{-}     & - 
\end{tabular}
}
\label{tab:rq4}
\vspace{-.18in}
\end{table}

\subsection{Scaling to Complex Programs (Q4)}\label{ss:rq4}

While TitanBench's tasks are already more complex than those in MBPP-DFY-153, they implement brute-force solutions that require minimal proof decomposition \eg less than two helper lemmas. To better demonstrate \systemname' capabilities on non-trivial verification tasks, we selected a set of challenging examples from the LeetCode repository and manually implemented\slash verified versions that go beyond simple brute-force approaches. Verifying these implementations requires non-trivial helper lemmas and detailed low-level reasoning \ie requiring significantly greater verification~effort.

The selected examples offer a balanced and meaningful range of complexity, with task difficulty determined by the number of verification lines and the number of helper lemmas required in the manually verified code. 
As shown in Table~\ref{tab:rq4}, the length of the verification code (LOVC) ranges from $21$ to $487$ lines, and the number of helper lemmas ranges from $1$ to $11$. Since the algorithms are more challenging to verify, we increase the generation timeout to $1500$ seconds. 
Because the complexity of these verification tasks surpasses the reasoning capacity of today's LLMs, we supply a proof outline to both the baseline and \systemname.\footnote{The textual proofs are available in the artifact.} These outlines ensure that the verification process begins from a sound and purposeful foundation, and they highlight how human insights can effectively complement LLM-based verification.
\smartparagraph{\systemname can prove algorithms that require $>$200 lines of verification code.} Looking at the results in Table~\ref{tab:rq4}, the baseline is able to verify only the two simplest tasks, while \systemname successfully verifies $7$ tasks within five runs. 
Interestingly, \systemname often provides more lemmas than the corresponding human-written solutions, meaning that the system tends to decompose proofs into smaller, more manageable steps, while humans can deal with complex reasoning within a single, more comprehensive lemma. In the unsolved Task \#8, \systemname is capable of solving partial proof obligations but fails to generate all of them. We note that \systemname starts to fail more often as the proof context grows in complexity. Nevertheless, the partially verified program can still be manually inspected by the user, who may complete the verification starting from the well-defined, partially correct proof.

\vspace{-.05in}
\section{Related Work}


\smartparagraph{Dafny and LLMs.} In the context of Dafny, recent work focuses mainly on generating the necessary annotations for verification~\cite{loughridge2024dafnybench,poesia2024dafnyannotator,mugnier2024laurel,wu2024automated}. 
%
Other studies aim to generate both the code body and its accompanying verification annotations through various strategies, including few-shot learning, Monte Carlo search, chain-of-thought prompting, retrieval-augmented generation, and feedback-driven techniques~\cite{misu2024towards,sun2024clover,mirchev2024assured,parnian2024pccc, brandfonbrener2024vermcts}. 
However, all these approaches are limited as they generate simple Dafny programs based on algorithmic descriptions. None of them refactors code or handle non-trivial generation tasks involving multiple lemmas, as with \systemname.
%

\smartparagraph{LLM-assisted verification in other programming languages.} 
Aside from Dafny, LLMs have been used for formal verification in other programming languages. Previous work explores C verification through tools such as VST~\cite{mukherjee2024towards} and Frama-C~\cite{kamath2024leveraging}, and Rust verification using Verus verifier~\cite{chen2024automated}. 
Lemur~\cite{wu2024lemur} formalizes the interaction between LLMs and verifiers as a sound proof calculus. 
Greiner \textit{et al.}~\cite{greiner2024automated} fine-tune LLMs to generate JML annotations, showing high syntactic validity and partial logical soundness of Java methods, but requires manual inspection for edge-case correctness.
SpecGen~\cite{ma2024specgen} uses LLMs with mutation-based refinement and conversational prompting to generate JML specifications, but has weak performance in the context of nested loops. None of these works have proposed to transiently refactoring the code to help LLMs for code verification.  
None of these works suggests using code refactoring to help LLMs in verifying code.

\smartparagraph{Autoformalization.} 
Some studies also utilize LLMs for autoformalization, \ie translating mathematical statements from natural language into formal specifications and proofs~\cite{wu2022autoformalization,zhou2024dont}. However, these systems focus solely on pure math problems (as opposed to programming) and they generate proofs in complex languages (\eg Lean or Isabelle), which are not as user-friendly as Dafny.

\smartparagraph{Refactoring in formal verification in non-LLM context}.  
Echo \cite{yin2009exploiting} showed that semantics-preserving refactorings, such as procedure splitting, loop rerolling, and reversing inlining, can simplify proofs by aligning specifications and reducing verification complexity. However, Echo is a framework that assists human users and does not perform decomposition automatically. For instance, when applying a decoupled loop transformation, an LLM must still verify that semantics are preserved independently. Echo’s evaluation on an optimized AES implementation was fully manual, and the work predates modern LLM-based verification.
In future work, we plan to integrate \systemname within a complementary semantics-preserving transformation framework, potentially enabling more effective and automated verification.

\section{Conclusions \& Future Work}
\label{sec:future}

In this paper, we identified key challenges to automate Dafny verification tasks using LLMs. We propose a novel curated dataset of non-trivial algorithms to evaluate LLMs' performance on complex Dafny verification tasks. We introduce \systemname, the first fully-automated system that decomposes code and proof, and integrates LLMs with advanced proof exploration and program repair techniques. \systemname outperforms baseline approaches and demonstrates its effectiveness in handling verification tasks that involve deeply nested loops and helper lemmas.

\smartparagraph{Future Work.} 
As discussed in Sec.~\ref{sec:evaluation}, manual check of the correctness of implementation and formal specification takes significant time. Further automating the process and increasing the size of our benchmark remains a future work.
We will also explore how to make \systemname generate more reusable lemmas that may reduce verifier solving time beyond their immediate proof context.
We will look into improving the code repair phase by using reinforcement learning to train\slash fine-tune an ad-hoc LLM capable of debugging Dafny errors.
%



\bibliographystyle{ACM-Reference-Format}
\bibliography{main}

\end{document}